\documentclass[sigconf]{aamas}  

\usepackage{booktabs}

\usepackage{algorithm} 
\usepackage[noend]{algpseudocode}
\usepackage{xcolor}
\usepackage{amsmath,bm}
\usepackage{subcaption}
\usepackage{graphicx}
\usepackage{xcolor}

\usepackage{flushend}

\begin{document}

\title{Fairness based Multi-Preference Resource Allocation \\ in Decentralised Open Markets}  



%
\author{Pankaj Mishra}
\affiliation{
  \institution{Nagoya Institute of Technology}
}
\email{m.pankaj.890@stn.nitech.ac.jp}
\author{Ahmed Moustafa}
\affiliation{
  \institution{Nagoya Institute of Technology}
}
\email{ahmed@nitech.ac.jp}
\author{Takayuki Ito}
\affiliation{
  \institution{Kyoto University}
}
\email{ito@i.kyoto-u.ac.jp }

\begin{abstract}

In this work, we focus on resource allocation in a decentralised open markets. In decentralised open markets consists of multiple vendors and multiple dynamically-arriving buyers, thus makes the market complex and dynamic. Because, in these markets, negotiations among vendors and buyers take place over multiple conflicting issues such as price, scalability, robustness, delay, etc. As a result, optimising the resource allocation in such open markets becomes directly dependent on two key decisions, which are; incorporating a different kind of buyers' preferences, and fairness based vendor elicitation strategy. Towards this end, in this work, we propose a three-step resource allocation approach that employs a reverse-auction paradigm. At the first step, priority label is attached to each bidding vendors based on the proposed priority mechanism. Then, at the second step, the preference score is calculated for all the different kinds of preferences of the buyers. Finally, at the third step, based on the priority label of the vendor and the preference score winner is determined. Finally, we compare the proposed approach with two state-of-the-art resource pricing and allocation strategies. The experimental results show that the proposed approach outperforms the other two resource allocation approaches in terms of the independent utilities of buyers and the overall utility of the open market.

\end{abstract}


\maketitle


\section{Introduction}

Real-world decentralised open markets such as supply chain management (SCM) \cite{yousefi2019supplier,wang2019interdependent}, cloud environments \cite{prasad2016combinatorial}, wireless sensors networks \cite{edalat2011combinatorial}, include multiple independent vendors and multiple independent buyers, with their conflicting objectives. In such environments, vendors offer various sets of resources to the dynamically arriving buyers. In this context, the existence of multiple independent vendors and multiple potential buyers, with their dynamic resource requirements, makes the environment highly complex and dynamic, which in turn leads to a vendor elicitation problem \cite{weber1998non}. In this context, dynamically arriving potential buyers aim to maximise their utilities by selecting a most appropriate vendor based on their different kinds of preferences. Therefore, there is a pressing need for a resource allocation approach that maximises the utilities of the buyers by optimally incorporating the different preferences of the buyers in such competitive and dynamic environments. In this context, several vendor elicitation strategies have been researched in the literature. For instance, Lee et al. \cite{lee2013real} proposed a cloud vendor elicitation approach in open cloud markets. However, their approach is based on considering only a single preference of the buyer, e.g., the price of different resources. Contrarily, in open markets, each buyer normally expresses multiple preferences, which might be conflicting in their turn, for example, cost and availability. Therefore, vendor elicitation based on multiple conflicting preferences should fairly accommodate the trade-offs amongst the conflicting preferences of potential buyers. In addition to the above discussions, it becomes clear auction-paradigm is the most appropriate choice for resource allocation in such decentralise open markets. In this regard, to employ an optimal and stable auction paradigm in competitive and dynamic markets, resource allocation approaches must be designed such that they maintain equilibrium in the market \cite{murillo2008fair}. To maintain equilibrium in such open market environments, three major challenges need to be addressed: (1) participation: encouraging the participation of vendors in the auction, (2) efficiency: the allocation of resources with optimal resource utilisation, (3) fairness: each vendor should be given a fair chance to win the auction. Therefore, to address the above-mentioned limitations, this research proposes a two-stage efficient resource allocation approach, which employs a reverse-auction paradigm in dynamic and decentralised open market environments. The proposed approach works in two steps as follows: 1) priority based fairness labelling for maintaining the equilibrium, 2) incorporating multi-preferences of the buyers,  and 3)winner determination. The contributions of this research are as follows:
\begin{itemize}
\item First, priority labelling for observing fairness.
\item Second, a multi-preference vendor elicitation algorithm is proposed, to aid potential buyers to select the best available vendors.
\item Third, an open market simulated environment is developed, that can simulate the dynamic arrival and departure of buyers with different resource requests.
\end{itemize}
{The rest of this paper is organised as follows. The problem formulation of multi-preference resource allocation in open market environments is introduced in Section 2. In Section 3, the experimental results are presented for evaluating the proposed approach. And in Section 4 related works are discussed. Finally, the paper is concluded in Section 5}

\section{Multi Preference Resource Allocation}

\begin{figure}[htp]
\vspace{.3in}
\centerline{\includegraphics[width=0.4\textwidth]{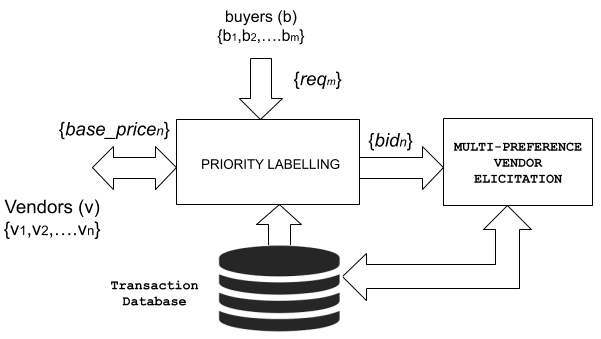}}
\vspace{.3in}
\caption{The proposed architecture of real-time resource allocation}
\label{fig:architecture}
\end{figure}

This section presents the architecture of the proposed Multi-Preference Resource Allocation (\textit{MPRA}) approach. Figure \ref{fig:architecture} depicts the architecture of the proposed two-stage resource allocation approach.

The architecture in Figure \ref{fig:architecture} consists of two modules which are; the \textit{PRIORITY LABELLING} module and the \textit{MULTI-PREFERENCE VENDOR ELICITATION} module. In specific, the proposed approach works as follows. First, a set of $J$ dynamically-arriving buyers submit their resource requests \{${req_1,\dots,req_j}$\} to the \textit{priority labelling} module. Then, based on these resource requests and the auction history from the transaction database, the \textit{PRIORITY LABELLING} module links a priority label to every bid (selling price) of the vendors. Then these priority labels along with the respective bid values of all the vendors are fed into the \textit{MULTI-PREFERENCE VENDOR ELICITATION} module; and finally, one vendor is elicited to satisfy each resource request. In this regard, the proposed two-stage resource allocation approach consists of two main stages, which are: (1) priority labelling, (2) final preference score and (3) winner determination. Both the stages are detailed in the following two subsections.


\subsection{Priority Labelling}

In this subsection, we discuss the proposed priority-based fairness mechanism for all the bidding vendors in the open markets. In this regard, a fairness mechanism aims to handle the primitive drop in bidders participating in the auction  \cite{murillo2008fair}. In specific, the fairness mechanism intends to avoid the monopoly of a single vendor in the market by allowing fair chances of winning to each bidder. In this regard, on receiving the bids from all the participating vendors in the auction, auctioneer attaches the priority label (\textit{pr}) to every bid from the vendors. In specific, on receiving the request $req_j=(rb_j,dl_j)$ from buyer $c_j$ and all the corresponding bids $bid_i^j=(rb_j,sp_i)$ from all the provider, where $i \in n$ and $j \in n$, priority label $pr_i$ is computed using Equation \ref{eq:priority} as follows.

\begin{equation}
  \label{eq:priority}
  pr=1-(loss/O)
\end{equation}

wherein, \textit{loss} is the number of times consumers lost in the last $O$ auctions, such that provider with $pr \in [0,1]$, $pr=0$ has highest priority and $pr=1$ has the lowest priority.

\subsection{Preference Scoring}
In this subsection, we discuss the proposed preference scoring mechanism that aims to enable the potential buyers to select a certain vendor, based on their multiple preferences. In this regard, different kinds of buyers preferences are of different metrics are to be scaled into a common scale of reference for comparison. In this context, the preference parameters of any buyer could be conflicting in nature, namely positive preference parameters or negative preference parameters. Specifically, if lower values of the preference parameters denote good characteristics, then the preference parameter is negative for example \textit{cost of the product}, otherwise, the preference parameter is positive. In this context, different kinds of preferences are normalised based on the simple additive weighting (SAW) mechanism \cite{zeng2003quality}. Using the \textit{SAW} mechanism, different values of different preference parameters are compared in two phases, which are; the scaling phase and the weighing phase. Further, we discuss each phase briefly as follows.

\subsubsection{Scaling Phase}
In this phase, the values of all the different preference parameters are scaled for further comparison. In this regard, scaling means normalising the values of all the different preference parameters that are submitted by the bidding vendors. In this context, each preference parameter for each bidding vendor is normalised based on Equations \ref{eq:pos} and \ref{eq:neg} for positive and negative preference parameters, respectively.

\begin{equation}
    \label{eq:pos}
  S_{iq}=\begin{cases}
    \frac{Q_q^{max}-Q_{iq}}{Q_{q}^{max}-Q_{q}^{min}}, & \text{if $Q_q^{max}-Q_q^{min}\neq 0$}.\\
    1, & \text{if $Q_q^{max}-Q_q^{min}= 0$}.
  \end{cases}
\end{equation}

\begin{equation}
    \label{eq:neg}
  S_{iq}=\begin{cases}
    \frac{Q_{iq}-Q_q^{min}}{Q_q^{max}-Q_{iq}^{min}}, & \text{if $Q_q^{max}-Q_q^{min}\neq 0$}.\\
    1, & \text{if $Q_q^{max}-Q_q^{min}= 0$}.
  \end{cases}
\end{equation}

where, $S_{iq}$ denotes the scaled value for vendor $i$ and preference parameter $q$, while $Q_{iq}$ denotes the value of preference parameter $q$, such that $q \in K$ and $i\in N$, for a market environment with $k$ preference parameters, and $n$ vendors.

\subsubsection{Weighing Phase}
In this phase, the scaled values of all the $k$ preference parameters are used to calculate the final preference vector $Q_i$, for each bidding vendor $i \in N$. The value of $Q_i$ is calculated using Equation \ref{eq:score}, 

\begin{equation}
  \label{eq:score}
 Q_i = \sum_{q=1}^{t}{S_{iq} \times P_q}
\end{equation}

where $P_q$ represents the preference weight of preference parameter $q$, which is decided independently by each buyer.

\subsection{Winner Determination}
In this subsection, we discuss the proposed winner determination strategy which is based on the previously computed priority labels and preference scores for each vendor. In specific, in this research, we aim at giving equal chance to all the bidding vendors, to avoid the monopoly in the auction paradigm and avoid bidder drop problem in the open market. In this work, we propose a winner determination algorithm, which is depicted in Algorithm \ref{algo:vea}. The proposed winner determination algorithm takes as input the multiple preferences of the buyer, along with the offered bids from all the bidding vendors. Further gives the winning vendor and the payment of the resources as output.

\begin{algorithm}
\caption{Winner Determination}
\label{algo:vea}
\begin{algorithmic}[1]
\State \textbf{Input:} \{bid(i,j)\}, where $i \equiv [1,N]$, $j \equiv [1,M]$
\State \textbf{Output:} $\{cost,winner\_vendor_j\}$ , $j \equiv [1,M]$
\For{buyer = 1 to $m$ \do} 
\For{vendor = 1 to $N$ \do} 
\State Compute $pr_{i}$, $i \in n$ and $q \in K$, using Equation \ref{eq:priority}
\For{criteria = 1 to $q$ \do} 
\State Compute $S_{iq}$, $i \in n$ and $q \in K$, Equation \ref{eq:pos},\ref{eq:neg}
\EndFor
\State Compute $Q_i$ , $i \in N$ using Equation \ref{eq:score} 
\EndFor
\State Sort all the bidders based on $pr$ in ascending order
\State \textit{priority\_list}: all the bidders with \textit{pr} below priority index ${p\_index}$
\State \textit{winning vendor} : max ($Q_k$), $k \in priority\_list$
\State \textit{cost}: average of all the vendors bid values in \textit{priority\_list}
\EndFor
\end{algorithmic}
\end{algorithm}

In the proposed winner determination algorithm, firstly all the bidders are sorted in ascending order of their priority label \textit{pr}. After that, bidders having priority label above priority index ${pr_{index}}$ \footnote{the value of priority index is set by the auctioneer at the beginning of an auction} are not further considered for resource allocation, whereas, rest of bidders are stored in a separate list named as $priority\_list$. Then finally the $Q$ values of all the bidders in the \textit{priority\_list} are compared among each other for resource allocation. Specifically, the bidder with a maximum $Q$ value is the winner and the buyer have to pay average bid value of all the bidders in \textit{priority\_list}. Finally, the Algorithm \ref{algo:dspra} depicts the proposed multi-preference resource allocation (\textit{MPRA}) approach. The proposed algorithm takes a set of vendors, their resource availability and all the requests of all the buyers. In the next section, we present the results of the extensive experiments that were conducted to evaluate the proposed resource allocation approach in a simulated open market environment.

\begin{algorithm}
\caption{Multi-Preference Resource Allocation (MPRA)}
\label{algo:dspra}
\begin{algorithmic}[1]
\State \textbf{Initialise:} set of N vendors :$\{1,\dots,N\}$, buffer $B$ 
\State \textbf{Initialise:}  set of resources $R_i \in N$, $\{R_1,\dots,R_N\}$
\State \textbf{Initialise:}  buffer for waiting buyers $B$, available vendors $A$
\State \textbf{Event:} Push requesting buyers $\{1,\dots,M\}$ in $B$
\State Checking for available vendors
\State \textbf{if} available
\State \hspace*{1em} Push available vendor in in buffer $A$
\State \textbf{if} set \textbf{A} is not empty
\State \hspace*{1em} Collect all the bid values of vendors
\State \textbf{else}
\State \hspace*{1em} Push the buyer's request in buffer $B$, wait for max wait time
\State \hspace*{1em} repeat \textbf{step} (5)
\State \textbf{End Event}
\State \textbf{Event:} Vendor becomes available
\State Allocate the available vendor to $A$ based on Algorithm \ref{algo:vea}
\State Check buyer in $B$
\State \textbf{if} $B$ is empty
\State \hspace*{1em} Repeat from \textbf{step}(5)
\State \textbf{else}
\State \hspace*{1em} Pop a buyer from the buffer $B$
\State \hspace*{1em} repeat from \textbf{step}(6)
\State \textbf{End Event}
\end{algorithmic}
\end{algorithm}

\section{EXPERIMENTAL SETUP AND RESULTS}

This section presents the results of the simulation experiments, that are conducted to evaluate the proposed resource allocation approach in an open market cloud environment. In order to simulate an open market cloud environment, we investigated three notable cloud simulators, namely; \textit{CLOUDSIM} \cite{calheiros2011cloudsim}, \textit{CLOUDSIM-PLUS} \cite{silva2017cloudsim} and \textit{GreenCloud} \cite{kliazovichpacket}. However, from our investigations, it became clear that, in their current stages of developments, these simulators do not provide native support for real-time cross-platform communications. Therefore, we choose to deploy a discrete-event simulation environment using the discrete-time python-based framework \textit{SimPy} \cite{matloff2008introduction}, to enable multiple cloud vendors to serve multiple dynamically arriving buyers. Towards this end, using this simulated cloud environment, we compare the proposed \textit{MPRA} approach with two other notable bidding-based resource allocation approaches, which are as follows: (1) the combinatorial double auction resource allocation approach (\textit{CDARA})  \cite{samimi2016combinatorial}: in this approach, the bid values represent the base prices of different resources. Then further, vendor elicitation is carried out based on the bid values that were set by vendors; (2) the indicator-based combinatorial auction-based approach (ICAA) \cite{kong2015auction}: in this approach, the bid values are calculated based on a mathematical function, that depends on the demands for different resources. Besides, vendor elicitation is also carried out based on the bid values that were set by vendors.

\subsection{Experimental Setup}

In this \textit{SimPy} based simulated cloud environment, each cloud service vendor is initialised with quantities of four types of resources; namely, computer processing speed (CPU), which is measured in millions of instructions per seconds (MIPS), memory, which is measured in megabytes (MB), storage,  which is also measured in megabytes (MB), and bandwidth (BW), which is measured in bits per seconds (B/S). In this regard, each cloud vendor is represented as $v \equiv \{AR,\{base\_price\},PQ\}$, where, \textit{AR} denotes the available set of resources, \textit{base\_price} denotes the base price for each resource type and \textit{PQ} denotes the set of preference parameters' values. On the other hand, each buyer is represented as $b \equiv \{RR, DL, P\}$, where \textit{RR} denotes the set of requested resources, \textit{DL} denotes the deadline for each resource request and \textit{P} represents the set of preference parameters' weights.
In addition, in these experiments, we consider three conflicting preference parameters, which are; (1) the cost of the requested resource; (2) the availability of each vendor and (3) the acceptance rate of each vendor; wherein, the cost of the requested resource is a negative preference parameter, while the other two parameters are positive preference parameters. Further, the evaluation of the proposed approach is concluded based on two different evaluation criteria, which are: (1) evaluation based on the performance of vendors; and (2) evaluation based on the performance of buyers. These three criteria are evaluated for all three approaches under the same experimental settings. Towards this end, we run the simulated cloud environment under four different experimental settings concerning the number of available vendors, i.e., with four, six, eight and twelve cloud-vendors (agents).  In addition, we run the three resource allocation approaches for 150 episodes (after convergence), each episode of length 2000 seconds, wherein 25 buyers arrive dynamically in each episode. We also set the value for  $pr_{index}$ to $0.4$. Further, we discuss the results of the three evaluation criteria in the following subsections.

\subsection{Evaluation Based on the Performance of Vendors}

In this subsection, the performance of all the bidding vendors is evaluated based on three parameters, which are: (1) \textit{average revenue}:- which is the average of all the revenues earned by each vendor during the 150 episodes. This parameter value is calculated for all the experimental settings and approaches, as shown in Table \ref{tab:avg_rev}; and (2) \textit{fairness}:- which is the ratio of the number of never-won vendors to the total number of vendors. This ratio indicates the active participation of vendors in the auction and its lower values depict the existence of bidder drop issue. This ratio is also calculated for all the experimental settings and approaches, as shown in Table \ref{tab:fair}. The results of the vendor performance evaluation based on these three parameters are interpreted as follows. Firstly, as shown in Table \ref{tab:avg_rev}, it is clear that the average revenue of the proposed \textit{MPRA} approach is approximately $45 \%$ higher as compared to the other two approaches. Meanwhile, as shown in Table \ref{tab:fair}, the \textit{MPRA} approach outperforms the other two approaches with higher fairness values in all the four experimental settings. Specifically, for the experimental setting with 12 vendors, fairness is significantly (60\%) higher as compared to the other two approaches. Therefore, from the above results, it becomes clear that the proposed \textit{MPRA} approach is capable of maximising the performance of the vendors in the open market cloud environment.

\begin{table} [!htb]
\caption{Revenue of Vendors}
\centering
\begin{tabular}{|c| c| c| c|} 

\hline
\# Vendors  & \textit{CDARA} & \textit{ICAA} & \textit{MPRA*}   \\ [0.5ex] 
\hline
4  & ${3.5*10^{7}}$ & ${4*10^{7}}$ & $\bm{5.5*10^{7}}$ \\ 
\hline

6  & ${3*10^{7}}$ & ${3.2*10^{7}}$ & $\bm{4.6*10^{7}}$ \\ 
\hline

8  & ${2.8*10^{7}}$ & ${2.8*10^{7}}$ & $\bm{4.2*10^{7}}$ \\ 
\hline

12  & ${2.8*10^{7}}$ & ${2.3*10^{7}}$ & $\bm{3.8*10^{7}}$ \\ 
\hline

\end{tabular}

\label{tab:avg_rev}
\end{table}

\begin{table} [!htb]
\caption{Fairness}
\centering
\begin{tabular}{|c| c| c| c|} 
\hline
\# Vendors  & \textit{CDARA} & \textit{ICAA} & \textit{MPRA*}   \\ [0.5ex] 
\hline
4  & 0.94 & 0.92 & $\bm{0.99}$ \\ 
\hline

6  & 0.82 & 0.88 & $\bm{0.97}$ \\ 
\hline

8  & 0.71 & 0.81 & $\bm{0.96}$ \\ 
\hline

12  & 0.48 & 0.53 & $\bm{0.82}$ \\ 
\hline

\end{tabular}

\label{tab:fair}
\end{table}

\subsection{Evaluation Based on the Performance of Buyers}

In this subsection, the performance of all the potential buyers is evaluated based on one parameter, which is: \textit{payment ratio}:- which is the average ratio of the prices that are paid by the buyers for their requested resources to the maximum prices that are offered by the bidding vendors for these resources. In this context, it represents the marginal difference between the paid price and the maximum offered price for any requested resource, as shown in Table \ref{tab:pmb}. As demonstrated in Table \ref{tab:pmb}, the payment ratio for the proposed \textit{MPRA} approach outperform the other two approaches. In this context, it is evident that the potential buyers which use the \textit{MPRA} approach pay marginally lesser prices for the requested resources as compared to the maximum offered prices by the bidding vendors. Therefore, from the above results, it is clear that the proposed approach is capable of maximising the performance of the buyers in the open market for the cloud environment.


\begin{table} [!htb]
\caption{Payment Ratio}
\centering
\begin{tabular}{|c| c| c| c|} 
\hline
\# Vendors  & \textit{CDARA} & \textit{ICAA} & \textit{MPRA*}   \\ [0.5ex] 
\hline
4 &  $0.82$ & $0.86$ & \textbf{0.71} \\ 
\hline
6 & $0.79$ & $0.92$ & \textbf{0.62} \\ 
\hline
8  & $0.78$ & $0.79$ & \textbf{0.58} \\
\hline
12 & $0.82$ & $0.78$ & \textbf{0.57} \\ 
\hline
\end{tabular}

\label{tab:pmb}
\end{table}


\section{Related Work}

To date, various approaches have been proposed to address resource allocation in decentralised open markets. In this regard, there is a consensus that efficient resource allocation approaches need to address one key challenge, that is, optimal winner determination strategy. To address this challenge, several approaches have been proposed. Specifically, an optimal strategy to which faces resource allocation in open market environments, that is the elicitation from the list of bidding vendors. In this regard, Kong et al. \cite{kong2015auction} and Prasad et el. \cite{prasad2016combinatorial} proposed a group of resource allocation approaches, where the buyer elicits the vendor with the minimum resource price (i.e., single preference parameter). Meanwhile, for multi-criteria decision making in web-service resource allocation, Zeng et al. \cite{zeng2004qos}, considered three quality of service attributes, namely, the execution-time, the reputation of vendors, and the reliability of buyers. Likewise for such multi-criteria decision-making situations, Afshar et al. \cite{afshari2010simple} proposed a SAW mechanism, which scales down all the different preference values and finally computes a single combined weight. Meanwhile, Pham et al. \cite{pham2015multi} and Mei et al. \cite{mei2015profit} considered multiple preferences for vendor elicitation. In specific, they investigated three preferences of the buyers, which are, the resource-price, the allocation-delay and the execution-speed. Therefore, in this work, with the consideration of the previous discussions, we chose to propose a reinforcement learning-based dynamic pricing policy and a SAW-based vendor elicitation strategy, to address the two previously mentioned challenges that face efficient resource allocation in dynamic and open market environments. 

\section{CONCLUSION}
This paper proposes a three-step fair resource allocation approach in dynamic decentralised markets. In the first step, the priority label is attached to vendors. Then, at the second step, the preference score is calculated for all the different kinds of preferences of the buyers. Finally, at the third step, based on the priority label of the vendor and the preference score winner is determined. Finally, we compare the proposed approach with two state-of-the-art resource pricing and allocation strategies. The experimental results show that the proposed approach outperforms the other two resource allocation approaches. As a result, the proposed approach enables both vendors and buyers to maximise their performances at the same time. The experimental results demonstrate the efficiency and fairness of the proposed resource allocation approach and its ability to maximise the overall performance of the open market environment. The future work is set to develop a mechanism that enables selective bidding by vendors based on multiple preferences.


\bibliographystyle{ACM-Reference-Format}  


\end{document}